\documentclass[reprint, amsmath,amssymb,showpacs,floatfix, aps, prb, twocolumn, superscriptaddress]{revtex4-1}
\usepackage{graphicx}
\usepackage{dcolumn}
\usepackage{bm}
\usepackage{color}
\usepackage{hyperref}
\hypersetup{colorlinks=true,citecolor=blue,linkcolor=blue, urlcolor=blue}
\newcolumntype{M}[1]{>{\centering\arraybackslash}m{#1}}
\newcolumntype{N}{@{}m{0pt}@{}}

\bibpunct{[}{]}{,}{n}{}{}

\def\nn{\nonumber}
\def\({\left(}
\def\){\right)}
\def\[{\left[}
\def\]{\right]}

\def\d{\delta}
\def\e{\epsilon}
\def\bser{\boldsymbol{\epsilon}_{r}}

\def\l{\lambda}
\def\w{\omega}

\def\kbt{k_BT}

\def\bfA{\mathbf{A}}
\def\bfB{\mathbf{B}}
\def\bfD{\mathbf{D}}
\def\bfE{\mathbf{E}}
\def\bfH{\mathbf{H}}
\def\bfQ{\mathbf{Q}}

\def\bfe{\mathbf{e}}
\def\bff{\mathbf{f}}
\def\bfg{\mathbf{g}}
\def\bfk{\mathbf{k}}
\def\bfp{\mathbf{p}}
\def\bfq{\mathbf{q}}
\def\bfr{\mathbf{r}}

\def\gs{\gamma_S}
\def\>{\rangle}
\def\<{\langle}
\def\barq{\bar{q}}

\def\kbt{k_BT}
\def\bsn{\boldsymbol{\nabla}}
\def\bsP{\boldsymbol{\Pi}}
\begin{document}
\title{\textsl{Ab Initio} Calculations of Exciton Radiative Lifetimes\\ in Bulk Crystals, Nanostructures and Molecules}

\author{Hsiao-Yi Chen}
\affiliation {Department of Applied Physics and Materials Science, 
California Institute of Technology, Pasadena, California 91125}
\affiliation {Department of Physics, 
California Institute of Technology, Pasadena, California 91125}

\author{Vatsal A. Jhalani}
\affiliation {Department of Applied Physics and Materials Science, 
California Institute of Technology, Pasadena, California 91125}

\author{Maurizia Palummo}
\affiliation {Dipartimento di Fisica and INFN, Universit$\grave{a}$ di Roma Tor Vergata, Via della Ricerca Scientifica 1, 00133 Roma, Italy}

\author{Marco Bernardi}
\email[E-mail: ]{bmarco@caltech.edu}
\affiliation {Department of Applied Physics and Materials Science, 
California Institute of Technology, Pasadena, California 91125}

\date{\today}

\begin{abstract}
\noindent  
Excitons are bound electron-hole pairs that dominate the optical response of semiconductors and insulators, especially in materials where the Coulomb interaction is weakly screened. 
Light absorption (including excitonic effects) has been studied extensively using first-principles calculations, but methods for computing radiative recombination and light emission are still being developed.   
Here we show a unified {\it ab initio} approach to compute exciton radiative recombination in materials ranging from bulk crystals to nanostructures and molecules. 
We derive the rate of exciton radiative recombination in bulk crystals, isolated systems, and in one- and two-dimensional materials, using Fermi's golden rule within the Bethe-Salpeter equation approach. 
We present benchmark calculations of radiative lifetimes in a GaAs crystal and in gas-phase organic molecules.  
Our work provides a general method for studying exciton recombination and light emission in bulk, nanostructured and molecular materials from first principles.
\end{abstract}
\maketitle
\section{Introduction \label{Introduction}}
\vspace{-10pt}
An exciton is a neutral excitation consisting of an electron-hole pair bound by the Coulomb interaction.  
In bulk metals, where the Coulomb interaction is screened by the conduction electrons, electron-hole pairs can be regarded as non-interacting, and excitons do not form. 
In semiconductors and insulators, and in particular in molecular and nanostructured materials, where electron-hole interactions are weakly screened, 
excitonic effects dominate the low-energy absorption spectrum and the radiative processes \cite{knox1963theory}. 
\\
\indent
Excitons play a key role in understanding carrier dynamics and accurately computing light emission processes in materials~\cite{park1966exciton,agranovich1995effect,schaller2005high,mikhnenko2015exciton}. 
Yet, calculations of radiative lifetimes typically employ simplified empirical models that can only qualitatively explain or fit the experimental data \cite{andreani1991radiative,ridley2013quantum, landsberg}, 
or are carried out in the independent-particle picture~\cite{delerue, Kioupakis2013, vandewalle}, neglecting excitons altogether. 
Over the last few years, first-principles approaches have been developed for accurately predicting exciton radiative lifetimes and light emission~\cite{spataru2005theory, 
palummo2015exciton, chen2018theory}.\\
\indent
These approaches employ the {\it ab initio} Bethe-Salpeter equation (BSE)~\cite{strinati1984effects,rohlfing2000electron} as a starting point to compute the exciton radiative lifetimes.  
A calculation of this kind was proposed by Spataru et al.~\cite{spataru2005theory} to compute the radiative lifetimes in a one-dimensional (1D) system (carbon nanotubes). 
We recently formulated the theory of exciton recombination and radiative lifetimes in two-dimensional (2D) materials~\cite{palummo2015exciton,chen2018theory}, 
where deriving the radiative lifetimes is more difficult and computing and diagonalizing the BSE Hamiltonian is more computationally demanding than in the 1D case. 
Our approach has enabled accurate predictions of the radiative lifetimes, as well as their temperature dependence and anisotropy, in novel 2D semiconductors~\cite{palummo2015exciton,chen2018theory}. 
However, for the main light emitters of technological interest, including bulk crystals, molecules, single quantum emitters, quantum dots and other zero-dimensional (0D) systems, 
an {\it ab initio} approach for computing exciton recombination and the associated radiative lifetimes has not yet been rigorously derived.
\\ 
\indent
In this work, we present a unified formulation of exciton radiative lifetimes in bulk crystals, 2D and 1D materials, and 0D isolated systems. 
The bulk and 0D cases are derived here from scratch, while the 2D and 1D cases, for which previous derivations exist, are briefly reviewed. 
We validate our approach by computing the radiative lifetimes in a GaAs crystal and in several gas phase organic molecules. 
Our work presents a broadly applicable approach to study light emission in materials, providing both the relevant equations and an {\it ab initio} 
workflow for computing radiative lifetimes in materials ranging from bulk crystals to nanostructures and molecules.
\\ 
\indent 
The manuscript is organized as follows. In Sec.~\ref{Theory} we briefly review the {\it ab initio} 
BSE approach and derive the second quantization of light in materials. 
In Sec.~\ref{sect:lifetimes} we present a general approach for computing exciton radiative lifetimes using Fermi's golden rule, 
and derive the radiative rate as a function of temperature for different cases, including bulk, 2D, 1D and 0D systems, each in a separate subsection. 
In Sec.~\ref{Sect:data} we present numerical calculations of radiative lifetimes in a GaAs crystal and in gas-phase organic molecules.
We summarize the results and discuss future research in Sec.~\ref{conclusion}.
%
%
\section{Theoretical Framework \label{Theory}}
\vspace{-10pt}
In this section, we briefly review the {\it ab initio} BSE approach \cite{strinati1984effects,rohlfing2000electron} for studying excitons from first principles (Sec.~\ref{subsect:Exciton/BSE}) 
and derive the second quantization of light in anisotropic materials (Sec.~\ref{subsect:Photon 2nd Quantization}). 
%
%
\subsection{Excitons and the Bethe-Salpeter Equation \label{subsect:Exciton/BSE}}
\begin{figure}[t]
\centering
\includegraphics[scale=0.22]{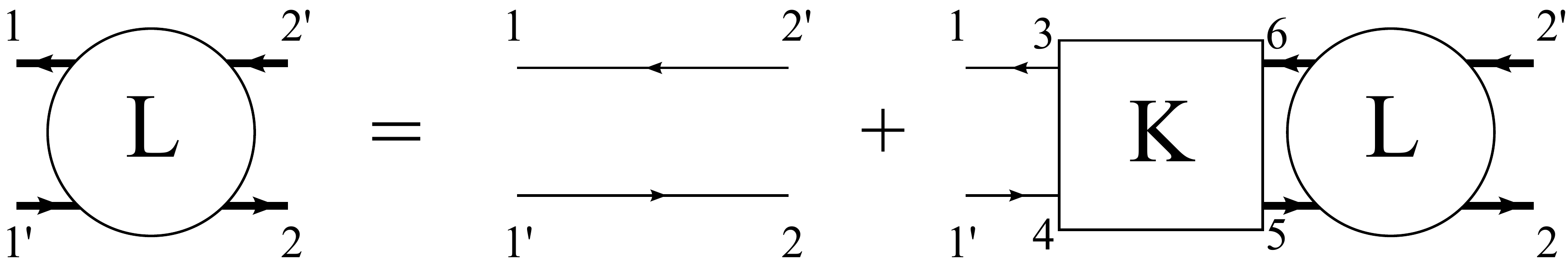}
\caption{The Bethe-Salpeter equation in its Dyson form, shown using Feynman diagrams. For details, see Ref.~\cite{strinati1984effects}.}
\label{Fig:BSE}
\end{figure}

\vspace{-10pt}
An exciton state can be represented in the so-called \lq\lq transition space\rq\rq~using pairs of electron-hole states as a basis. 
In a periodic system, these states are Bloch wavefunctions characterized by the band index and crystal momentum. 
Within the Tamm-Dancoff approximation, which ignores antiresonant transition terms \cite{fetter2012quantum}, 
an exciton state $S$ with center-of-mass momentum $\bfQ$ can be written as a superposition of electron-hole states,
\begin{equation}
\label{eq-TD}
|S\bfQ\>=\sum_{vc\bfk}A^{S\bfQ}_{vc\bfk}|v\bfk\>_h|c\bfk+\bfQ\>_e \,\,,
\end{equation}
where $v$ labels the valence and $c$ the conduction bands, $\bfk$ is the electron crystal momentum, and the subscripts $e$ and $h$ denote electron and hole states, respectively. 
The expansion coefficients $A^{S\bfQ}_{vc\bfk}$ can be obtained by solving the BSE, which is shown diagrammatically in Fig.~\ref{Fig:BSE} and can be written as~\cite{strinati1984effects}
\begin{equation}
\label{eq-BSE}
L(12;1'2')=L_0(12;1'2')+L_0(1\bar{4};1'\bar{3}) K(\bar{3}\bar{5};\bar{4}\bar{6})L(\bar{6}2;\bar{5}2'),
\end{equation}
where we use numbers for spacetime coordinates, i.e., $1=(\bfr_1,t_1)$, and the overlines denote dummy integration variables. 
Here, $L(12;1'2')$ is the exciton correlation function and $L_0(12;1'2')=G_1(1,2')G_1(2,1')$ its non-interacting counterpart, with $G_1$ the one-body Green's function. 
The key ingredient in the BSE is the kernel $K(\bar{3}\bar{5};\bar{4}\bar{6})$, which encodes the interaction between the electron and hole. Within the GW approximation, it can be written as
\begin{eqnarray}
K(35;46)&=&-i \d(3,4)\d(5^-,6)v_c(3,6)\nn\\&&
+i \d(3,6)\d(4,5)W(3^+,4),
\end{eqnarray}
where the first term is the exchange and the second  the screened Coulomb interaction. 
\\
\indent
In the transition basis defined in Eq.~(\ref{eq-TD}), solving the BSE reduces to the eigenvalue problem \cite{rohlfing2000electron}
\begin{equation}
\label{eq-BSEnum}
(E_{c\bfk+\bfQ}-E_{v\bfk})A^{S\bfQ}_{vc\bfk}+\sum_{v'c'\bfk'}K_{vc\bfk,v'c'\bfk'}A^{S\bfQ}_{v'c'\bfk'}=E_{S}(\bfQ)A^{S\bfQ}_{vc\bfk}
\end{equation}
where $E_{c\bfk+\bfQ}$ and $E_{v\bfk}$ are the electron and hole quasiparticle energies, and the kernel $K_{vc\bfk,v'c'\bfk'}$ can be written in the electron-hole basis as \cite{rohlfing2000electron}:
\begin{eqnarray}
&&K_{vc\bfk,v'c'\bfk'}=i \psi_{v\bfk}\(\bar{4}\)\psi^*_{c\bfk+\bfQ}\(\bar{3}\)
\nn\\&&\qquad\qquad\qquad\times K\(\bar{3}\bar{5},\bar{4}\bar{6}\)
\psi^*_{v'\bfk'}\(\bar{5}\)\psi_{c'\bfk'+\bfQ}\(\bar{6}\),
\end{eqnarray}
where $\psi_{c(v)\bfk}$ are conduction (valence) single-electron Bloch wavefunctions.
\\
\indent
In practice, the {\it ab initio} BSE is solved by constructing the kernel (typically from the static RPA dielectric function) and diagonalizing Eq.~(\ref{eq-BSEnum}) with a linear algebra package. Several codes implement this workflow, including {\sc Yambo} \cite{marini2009yambo}, {\sc Abinit} \cite{gonze2016recent} and {\sc BerkeleyGW} \cite{deslippe2012berkeleygw}.
\vspace{20pt}
%
%
\subsection{Quantization of Light in Bulk Materials}
\label{subsect:Photon 2nd Quantization}
\vspace{-10pt}
To derive the radiative lifetime in bulk crystals with general symmetry, we present the non-relativistic theory of second quantization of light in bulk materials. 
The theory for isotropic bulk materials is presented in Ref.~\citep{glauber1991quantum}, and extended here to anisotropic bulk materials. 
We write the dielectric tensor in diagonal form, $\bser=$ diag$(\e_x,\e_y,\e_z)$, and work in a generalized Coulomb gauge in which $\bsn\cdot\left(\e_0\bser \bfE \right) =0$. The equation of motion for the vector potential $\bfA$ becomes
\begin{equation}
\label{eq-EOM_A1}
-\mu_0\e_0\bser\frac{\partial^2 
\bfA}{\partial^2 t}=\bsn\times(\bsn \times \bfA)=\bsn(\bsn\cdot \bfA)-\bsn^2\bf{A} \,,
\end{equation}
using which we can construct the Lagrangian
\begin{eqnarray}
\mathcal{L}=
\frac{1}{2}\int d\bfr \[\e_0\dot { \bfA} ^T(\bfr)\bser\dot { \bfA}(\bfr)-\frac{\(\bsn\times\bfA\)^2}{\mu_0}\].
\end{eqnarray}
Since the conjugate momentum is $\bsP (\bfr)=\e_0\bser\dot{\bfA}(\bfr)$, the Hamiltonian reads 
\begin{equation}
\label{eq-Hamiltonian}
\mathcal{H}=\int \!\!d\bfr\, \bsP\dot{\bfA}-\mathcal{L}=\frac{1}{2}\int d\bfr
\[\frac{\bsP^T \bser^{-1}\bsP}{\e_0}+\frac{\(\bsn\times\bfA\)^2}{\mu_0}\].
\end{equation}
%
To define creation and annihilation operators for second quantization, we solve Eq.~(\ref{eq-EOM_A1}) and obtain
\begin{equation}
\label{Eq:vector potential}
\bfA=\sum_{\l\bfq}\sqrt{\frac{\hbar}{2V\w_{\l\bfq}\e_0}}
\(\hat{a}_{\l\bfq}\bfe_{\l\bfq}e^{i\(\bfq\cdot\bfr+\w_{\l\bfq}t\)}+h.\,c.\),
\end{equation}
where $h.\,c.$ stands for Hermitian conjugate, and the photon frequencies $\omega_{\l\bfq}$ and polarization vectors $\bfe_{\l\bfq}$ (where $\l$ labels the mode, and $\bfq$ the photon wavevector) are obtained by solving Eq.~(\ref{photon dispersion}) in Appendix \ref{Appen:photon quant}, 
with the polarization vectors satisfying the generalized orthogonality condition~\cite{Kumar2002}
\begin{equation}
\label{Eq:ortho}
\bfe^\dagger_{\l \bfq}\bser\bfe_{\l' \bfq}=\delta_{\l,\l'}.
\end{equation}
This equation, together with eq.~(\ref{photon dispersion}) in Appendix \ref{Appen:photon quant}, 
provide the photon frequencies and polarization vectors needed to compute the radiative lifetime in bulk crystals of any given symmetry. 
Using these results, the electromagnetic field Hamiltonian in Eq.~(\ref{eq-Hamiltonian}) can be converted to the standard quantum oscillator form, 
$\mathcal{H}=\sum_{\l\bfq}\hbar\w_{\l\bfq}\left(\hat{a}^\dagger_{\l\bfq}\hat{a}_{\l\bfq}+1/2\right)$. 
Additional details are provided in Appendix \ref{Appen:photon quant}. 
\vspace{50pt}
%
%
\section{EXCITON RADIATIVE LIFETIMES \label{sect:lifetimes}}
\subsection{General theory\label{sect:general theory}}
\begin{figure*}[t]
\includegraphics[scale=0.7]{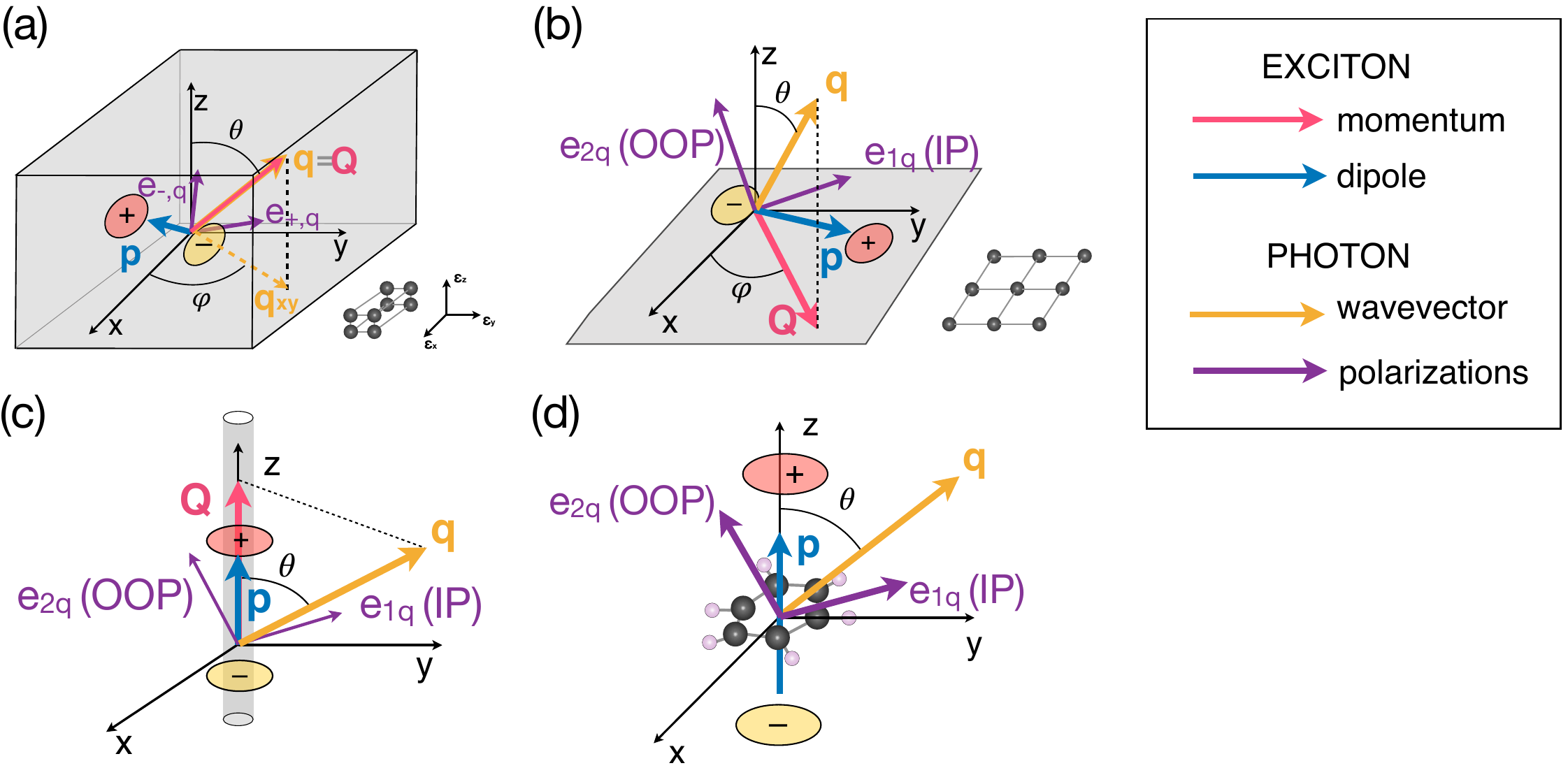}
\caption{Schematic of the exciton and photon quantities employed in this work. Each panel corresponds to a different dimensionality.   
(a) Bulk (three-dimensional) anisotropic material, in which momentum conservation requires $\bfq=\bfQ$, and the photon polarizations are nondegenerate and specified by the solution of the Maxwell equations [see Eq.~(\ref{photon dispersion}) in Appendix~\ref{Appen:photon quant}]. 
(b) Two-dimensional material, in which the exciton transition dipole $\bfp_S$ lies in the $xy$ plane containing the material, and the in-plane projection of the emitted photon wavevector equals the exciton momentum, namely $\bfQ=(\bfq\cdot \hat{Q})\hat{Q}$. 
(c) One-dimensional material, where both the exciton momentum and transition dipole lie along the material direction $z$, and momentum conservation imposes $\bfQ=\bfq\cdot \hat{z}$. 
(d) Isolated (zero-dimensional) system, with no constraints on the exciton momentum, photon wavevector and transition dipoles. 
In all cases, when the two photon polarizations are degenerate, the polarization vectors $\bfe_{\l\bfq}$ are chosen as in-plane (IP) and out-of-plane (OOP), where the IP component is in the $xy$ plane and the OOP in the $\bfq-\hat{z}$ plane.}
\label{Fig:coordinate}
\end{figure*}

\vspace{-10pt}
We use the minimal coupling Hamiltonian to describe the interaction between electrons and photons, $H_{\rm int}=-\frac{e}{m} \bfA\cdot \bfp$, 
where $\bfp$ is the momentum operator and $\bfA$ the vector potential in second quantized form (here and below, $e$ and $m$ are the electron charge and mass, respectively, and we use SI units)~\cite{loudon2000quantum}. 
The radiative recombination rate at zero temperature for an exciton in state $S$ with center-of-mass momentum $\bfQ$ can be written using Fermi's golden rule as
\begin{eqnarray}
\label{eq Fermi Golden}
&&\gamma_S(\bfQ)=
\frac{2\pi}{\hbar}
\sum_{\l\bfq}
\left|\langle G,1_{\l\bfq}|H_{\rm int}|S\bfQ,0\rangle\right|^2
\delta\( E_{S}(\bfQ)-\hbar \w_{\l\bfq}\)\nn\\
&&~~=
\frac{\pi e^2}{\e_0m^2V}
\sum_{\l\bfq}
\frac{1}{\w_{\l\bfq}}
\left|\bfe_{ \l\bfq}\cdot\bfp_S(\bfQ)\right|^2
\delta\( E_{S}(\bfQ)-\hbar \w_{\l\bfq}\)\nn,\\
\end{eqnarray}
where the initial state $|S\bfQ,0\rangle$ consists of an exciton and zero photons, and the final state $|G,1_{\l\bfq}\>$ is the electronic ground state plus one emitted photon with polarization $\lambda$ and wavevector $\bfq$; $V$ is the volume of the system.   
The summation runs over the two photon polarizations and all possible wavevectors $\bfq$ of the emitted photon, which has energy $\hbar\omega_{\l\bfq}$, while the delta function imposes energy conservation. 
The transition dipole $\bfp_S(\bfQ)=\langle G|\bfp|S\bfQ\rangle$ is in general a vector with complex-valued components (in 2D and 1D systems, the only nonzero components are those in the plane or line containing the material, respectively). In practice, we use the velocity operator and compute the transition dipole as $\bfp_S(\bfQ) = (-im/\hbar)\<G|[\mathbf{x}, H_{\rm KS}]|S\bfQ\>$ to correctly include the nonlocal part of the Kohn-Sham Hamiltonian, $H_{\rm KS}$~\cite{sangalli2017optical}. For light emission, the values of $\bfQ$ compatible with energy conservation are very small. 
For this reason, we approximate the dipole of an exciton $|S\bfQ\rangle$ as $\bfp_S(\bfQ) \approx \bfp_S(0)$ by solving, as is standard, the BSE at $\bfQ = 0$.
\\
\indent
The radiative lifetime at finite temperature $T$ for a given exciton state $S$ can be computed by assuming that the exciton momentum $\bfQ$ has a thermal equilibrium distribution, 
which is a good approximation when (as is common) the thermalization process is much faster than radiative recombination~\cite{wolfe1982thermodynamics}. 
We can thus write the radiative rate of the exciton state $S$ as the thermal average
\begin{equation}
\label{Eq:gammaST}
\<\gamma_S\>(T)=\frac{\int\! d\bfQ\,\, e^{-E_S(\bfQ)/k_BT}\, \gamma_S(\bfQ)} {\int \! d\bfQ\, e^{-E_S(\bfQ)/k_BT}}.
\end{equation}
The radiative lifetime is defined as the inverse of the radiative rate, $\<\tau_S\> = \<\gamma_S\>^{-1}$.
We employ an isotropic effective mass approximation for the exciton dispersion~\footnote{The parabolic exciton dispersion assumed here may be inadequate for Frenkel excitons. Since finite center-of-mass momentum BSE calculations are becoming available, rather than assuming a model it will ultimately be preferable to directly use a computed exciton dispersion in our lifetime formulas. The Wannier or Frenkel character of the exciton becomes less relevant if one uses a computed exciton dispersion or effective mass rather than a model.},
\begin{equation}
\label{Eq:parabola dispersion}
E_S(\bfQ)=E_S(0)+\frac{\hbar^2 Q^2}{2M_S},
\end{equation}
where the exciton mass $M_S$ is approximated as the sum of the electron and hole effective masses, $M_S=m_{e}^*+m_{h}^*$.\\
\indent
Note that the exciton dispersion and effective mass tensor can also be computed (rather than assumed) by solving the BSE with a finite exciton momentum \cite{gatti2013exciton,qiu2015nonanalyticity};  
this is particularly important in those cases in which a non-parabolic exciton dispersion is expected. 
For example, Cudazzo et al.~\cite{Cudazzo-2016} have shown that in 2D materials the exciton dispersion can be either linear or parabolic, depending on the character of the exciton wavefunction at finite 
$\mathbf{Q}$, and Qiu et al.~\cite{qiu2015nonanalyticity} have shown that of the two lowest-energy bright excitons in MoS$_2$, on has a linear and the other a parabolic dispersion. Here we focus on computing the radiative lifetime for excitons with a parabolic dispersion, and show in Appendix~\ref{Appen:non-analytic} the corresponding results for excitons with a linear dispersion.
In the following, we will also assume that the exciton mass is large enough for us to set in the delta functions $E_S(\bfQ)-\hbar\w_{\l\bfQ}\approx E_S(0)-\hbar\w_{\l\bfQ}$. 
\\
\indent
When only the lowest-energy bright exciton contributes to the photoluminescence, Eq.~(\ref{Eq:gammaST}) is a good approximation for the radiative rate. 
When multiple exciton states are occupied, an additional average is needed to include the contributions from all occupied exciton states. 
Assuming that the exciton states are occupied according to a thermal equilibrium distribution, the effective radiative rate one expects to observe experimentally is:  
\begin{equation}
\label{Eq:3D gamma eff} 
\<\gamma(T)\>_{\rm eff}=\frac{\sum_S \<\gamma_S\> e^{-E_S(0)/k_BT}}{\sum_S e^{-E_S(0)/k_BT}}. 
\end{equation} 
\\
\indent
Below, we derive the exciton radiative recombination rate as a function of temperature in materials with different dimensionality. The key quantities employed in the derivations, including the coordinates, the exciton momentum $\bfQ$ and transition dipole $\bfp_S$, and the photon polarization vectors $\bfe_{\lambda \bfq}$, are shown schematically in Fig.~\ref{Fig:coordinate} for each case discussed below. The equations for the bulk and 0D cases are derived here from scratch, 
while the 2D and 1D cases, which have been previously investigated, are reviewed briefly for completeness.
%
%
\subsection{Bulk (3D) materials \label{sect:bulktheory}}
\vspace{-10pt}
We consider a non-magnetic and non-absorbing~\footnote{For the sake of studying light emission, this assumption has negligible effects as it amounts to neglecting re-absorption or other dynamical processes of the emitted photons.} anisotropic bulk crystal, in which the static (zero-frequency) dielectric tensor can be written as
\begin{equation}
\bser={\rm diag}(\e_{x},\e_{y},\e_z).
\end{equation}
In crystals with cubic, tetragonal, orthorhombic and hexagonal symmetry, we orient the crystallographic axes along the $\{x,y,z\}$ cartesian directions,  
and in the uniaxial (tetragonal and hexagonal) cases we additionally orient the principal axis along the $z$ direction. 
In crystal classes with lower symmetry, including monoclinic and triclinic, we orient the principal axes (i.e., the eigenvectors of $\bser$) along the cartesian directions. 
With these choices, our treatment is general and can account for any crystal symmetry~\cite{newnham}. 
The photon energy in such an anisotropic material is modified according to the dielectric tensor. For a given photon wavevector $\bfq=(q_x,q_y,q_z)$, there are two propagating modes as solutions to Maxwell's equations; they correspond to the two photon polarizations~\cite{glauber1991quantum}, and their frequencies $\w_{\pm\bfq}$ are the solutions of Eq.~(\ref{photon dispersion}) in Appendix \ref{Appen:photon quant}:
\begin{equation}
\w_{\pm\bfq}^2=\frac{-\(\frac{\barq_x^2}{\e_x}+\frac{\barq_y^2}{\e_y}+\frac{\barq_z^2}{\e_z}\)\pm\tilde{\w}_{\bfq}^2}{2\mu_0\e_0},
\label{eq-modes}
\end{equation}
with
\begin{equation}
\tilde{\w}_{\bfq}^2=\sqrt{\(\sum_{\alpha}\frac{\barq_\alpha^2}{\e_\alpha}\)^2-4q^2\sum_\alpha\frac{q_\alpha^2\e_\alpha}{\e_x\e_y\e_z}}\,\,,
\end{equation} 
where $\alpha$ denotes the cartesian coordinates $\{x,y,z\}$, and $\barq_\alpha^2=q_\alpha^2-q^2$. 
The corresponding polarization vectors for the two modes are
\begin{equation}
\label{Eq:polarization}
\bfe_{\pm \bfq}=
\frac{1}{\Lambda_q}
\begin{pmatrix}
\vspace{3pt}
q_x(\w_{\pm\bfq}^2 \, \mu_0\e_0\e_x-q^2)\\
\vspace{3pt}
q_y(\w_{\pm\bfq}^2\,\mu_0\e_0\e_y-q^2)\\
q_z(\w_{\pm\bfq}^2\,\mu_0\e_0\e_z-q^2)
\end{pmatrix}
\end{equation} 
up to a normalization constant $\Lambda_q$; for details, see Appendix~\ref{Appen:photon quant}.  
This solution applies to photons propagating in anisotropic materials with $\e_x\neq \e_y \neq \e_z$. 
For materials with axial or cubic symmetry, in which, respectively, two or three of the diagonal components of the macroscopic dielectric tensor are equal, 
the frequencies and polarization vectors have simpler expressions, which can be derived from the general case discussed here.  
\\
\indent
For an exciton in state $|S\bfQ\>$ with momentum $\bfQ=(Q_x,Q_y,Q_z)$, we obtain the radiative recombination rate by applying Fermi's Golden rule [see Eq.~(\ref{eq Fermi Golden})].   
Momentum conservation fixes the emitted photon wavevector to $\bfq=\bfQ$ [see Fig.~\ref{Fig:coordinate}(a)], and the summation over $\l$ adds together the contributions from the $\w_{\pm\bfq}$ solutions. 
As mentioned before, we approximate the transition dipole by evaluating it at $\bfQ=0$, 
\begin{equation}
\label{Eq:3D-dipole}
\langle G|\bfp|S(\bfQ)\rangle\approx\langle G|\bfp|S(0)\rangle=p_{Sx}\hat{\bf x}+p_{Sy}\hat{\bf y}+p_{Sz}\hat{\bf z} ,
\end{equation}
with complex components $p_{S\alpha}$. 
Using these results, the exciton radiative recombination rate at zero temperature becomes 
\begin{eqnarray}
&&\gamma_S^{\rm 3D}(\bfQ)=
\frac{\pi e^2}{\e_0m^2V}\nn\\
&&~\times
\sum_{\l=\pm}
\left|\sum_{\alpha} \frac{p_{S\alpha}\,q_\alpha(\w^2_{\l\bfQ} \mu_0\e_0\e_\alpha-q^2)}
{\Lambda_q}
\right|^2
\frac{\delta\(E_S(\bfQ)-\hbar\w_{\l \bfQ}\)}
{\w_{\l \bfQ}}.\nn\\
\label{eq-gamma3d}
\end{eqnarray}

Next, we specialize our discussion to cubic or isotropic materials with a dielectric constant $\e$ [i.e., with dielectric tensor $\bser={\rm diag}(\e,\e,\e)$]. 
Radiative lifetime calculations for an anisotropic bulk crystal will be presented elsewhere. 
Due to symmetry, in the cubic or isotropic case the two modes in Eq.~(\ref{eq-modes}) become degenerate, with polarization vectors perpendicular to each other and to the direction of photon propagation. 
Following a convention we recently used in the 2D case \cite{chen2018theory}, we orient one of the two polarization vectors to lie in the $xy$ plane, and call this vector \lq\lq in-plane'' (IP). 
The other polarization vector then has a nonzero $z$ component, and is called \lq\lq out-of-plane'' (OOP). These two polarization vectors can be written in spherical coordinates as
\begin{eqnarray}
\label{Eq:IP/OOP}
{\rm IP}&:&\bfe_{1\bfq}=\frac{1}{\sqrt{\e}}(-\sin\varphi,\cos\varphi,0)\nn\\
{\rm OOP}&:&\bfe_{2\bfq}=\frac{1}{\sqrt{\e}}(-\cos\theta\cos\varphi,-\cos\theta\sin\varphi,\sin\theta),\nn\\
\end{eqnarray}
where $\theta$ is the polar and $\varphi$ the azimuth angle of the photon wavevector $\bfq$~[see Fig.~\ref{Fig:coordinate}(a)]. 
Substituting in Eq.~(\ref{eq-gamma3d}), we obtain the radiative rate at zero temperature for cubic or isotropic bulk materials (see Appendix \ref{Appen:3D}):
\begin{eqnarray}
\label{Eq:iso rate}
&&\gamma_{S}^{\rm 3D, iso}(\bfQ)=
\frac{\pi e^2}{\e_0m^2Vc\,Q\sqrt{\e}}
\left\{
\left|\frac{p_{Sx}Q_y-p_{Sy}Q_x}{Q_{xy}}\right|^2_{\rm IP}
\right.\nn\\
&&\left.
+
\left|
\frac{Q_xp_{Sx}+Q_yp_{Sy}}{Q_{xy}}
\frac{Q_{z}}{Q}
-
p_{Sz}
\frac{Q_{xy}}{Q}
\right|_{\rm OOP}^2
\!\right\}
\delta\!\(\!E_S(Q)-\frac{\hbar c Q}{\sqrt{\e}}\!\)\!\!.
\nn\\
\end{eqnarray}
This equation correctly accounts for momentum conservation, which in the bulk case translates to the condition $\mathbf{q}=\mathbf{Q}$. 
Different from the 0D, 1D and 2D cases, this result implies that excitons with zero center of mass momentum cannot decay radiatively, 
since the resulting photon would possess zero momentum and energy. 
In contrast, the equations employed so far to compute from first principles radiative decay in bulk employ either simplified independent-particle approaches~\cite{Kioupakis2013,vandewalle}, which neglect excitons altogether, or formulas appropriate for 0D or 1D~\cite{hu2018photocatalytic}, which fail to take into account momentum conservation. Note that neglecting excitons does not merely change the transition dipole and energies, but rather, the exciton momentum and dispersion are essential to accurately computing the temperature and polarization dependence of light emission.\\
\indent 
The radiative recombination rate of a given exciton state $S$ at temperature $T$, for isotropic bulk crystals under the assumption that the exciton momentum has a thermal equilibrium distribution, 
is obtained using Eq.~(\ref{Eq:gammaST}) as (see Appendix \ref{Appen:3D})
\begin{eqnarray}
\label{Eq:gammaT}
\<\gamma_S^{\rm 3D, iso}\>(T)=
\frac{8\sqrt{\pi \e}\,e^2\,\hbar \,p_S^2}{3\e_0 m^2VE_S(0)^2}
\(\frac{E_S(0)^2}{2M_Sc^2k_BT}\)^{3/2},
\end{eqnarray}
where the exciton energy $E_S(0)$ and the transition dipole $\bfp_S$ (and $p_S^2=|\bfp_S|^2$) are obtained by solving the BSE. 
The $T^{-3/2}$ temperature dependence of the radiative rate (and thus, the $T^{3/2}$ temperature dependence of the radiative lifetime) is consistent with previous semiempirical theoretical treatments~\cite{Lasher1964} and with low-temperature experimental data \cite{im1997radiative}.\\
\indent 
For bulk crystals with a low exciton binding energy ($<0.1$ eV), additional thermal effects include exciton dissociation and equilibration with free carriers \cite{wolfe1982thermodynamics}. 
This topic has been studied extensively experimentally; 
the net effect of the coexistence between excitons and carriers is an increase in the radiative lifetime, which can be important near room temperature and can cause the radiative lifetime to deviate significantly from the $T^{3/2}$ trend~\cite{im1997radiative}. 
Such coupled exciton-carrier dynamics can be treated with kinetic models but is still beyond the reach of first-principles calculations.
\\
\subsection{Two-dimensional materials}
\vspace{-10pt}
Novel 2D semiconductors, such as transition metal dichalcogenides and related layered materials, exhibit unique optical properties and strongly bound excitons that govern their light absorption and emission~\cite{Bernardi-review}. 
We have recently proposed a first-principles approach to compute the radiative lifetime in such 2D materials, as well as its angular and polarization dependence, 
which gives rise to anisotropic light emission~\cite{palummo2015exciton,chen2018theory}. 
In our approach, exciton recombination is still described using the Fermi Golden rule in Eq.~(\ref{eq Fermi Golden}). However, due to the lower dimensionality, the transition dipole is restricted to the 2D plane containing the material:
\begin{equation}
\bfp_S=p_{Sx}\hat{\bf x}+p_{Sy}\hat{\bf y},
\end{equation}
with complex components $p_{Sx}$ and $p_{Sy}$. Furthermore, since translation symmetry applies only in the plane containing the material, momentum conservation is imposed on the in-plane projection of the emitted photon wavevector, using $(\bfq \cdot \hat{Q})\hat{Q}=\bfQ$ [see Fig.~\ref{Fig:coordinate}(b)]. 
Unlike the bulk case, photons are emitted into the vacuum surrounding the 2D material (unless a substrate is present), and thus the emitted photons exhibit two degenerate polarizations. 
Following the same convention as in the isotropic bulk case, the IP and OOP polarizations are chosen as in Eq.~(\ref{Eq:IP/OOP}) with $\e=1$. 
Upon integrating over all final photon states, we obtain the radiative recombination rate of an exciton $S$ with momentum $\bfQ$ in a 2D material at zero temperature~\cite{chen2018theory}:
\begin{widetext}
\begin{equation}
\label{gamma0}
\gamma^{\rm 2D}_S(\bfQ) = \gamma^{\rm 2D}_S(0)\cdot
\left(\frac{E_S(0)}{\sqrt{E_S^2(Q)-\hbar^2c^2Q^2}}\right) \left\{
\left|- \frac{ p_{Sx}}{p_S}\sin\varphi+\frac{ p_{Sy}}{p_S}\cos\varphi \right|^2_{\rm IP} + 
\frac{E_S(Q)^2-\hbar^2c^2Q^2}{E_S(Q)^2}\left|\frac{ p_{Sx}}{p_S}\cos\varphi +\frac{ p_{Sy}}{p_S}\sin\varphi \right|_{\rm OOP}^2
\right\}
\end{equation}
\end{widetext}
where $\gamma^{2D}_S(0)=\frac{e^2 p_S^2}{\e_0 m^2c A E_S(0)}$ is the recombination rate for $\bfQ=0$ and $A$ is the
area of the system in the $xy$ plane.  
Note that due to momentum conservation there is an upper limit of $Q_0$ to the momentum of an exciton that can recombine radiatively; this limit occurs when a photon is emitted in the plane of the material, in which case $E_S(\bfQ) = \hbar c Q_0$. 
Excitons with momentum $Q>Q_0$ cannot emit a photon, and their radiative recombination rate vanishes since energy and momentum cannot be simultaneously conserved upon photon emission.\\
\indent
At finite temperature $T$, the exciton radiative lifetime can be computed by assuming, similar to the bulk case, a parabolic exciton dispersion 
$
E_S(\bfQ)=E_S(0)+\frac{\hbar^2Q^2}{2M_S}$,
where $M_S$ is an in-plane isotropic exciton effective mass. Taking the thermal average in Eq.~(\ref{Eq:gammaST}) of the 2D radiative rate in Eq.~(\ref{gamma0}), we obtain the radiative lifetime~\cite{palummo2015exciton}
\begin{equation}
\label{Eq:gammaT2D}
\<\tau_S^{\rm 2D}\> (T) 
=
\gamma_S^{\rm 2D}(0)^{-1}
\times
\frac{3}{4}
\left(\frac{2M_Sc^2k_BT}{E_S(0)^2}\right).
\end{equation}
This formula has been applied in our recent work \cite{palummo2015exciton}, giving temperature dependent radiative lifetimes in excellent agreement (within 5$-$10~\%) 
with experimental results obtained by transient photoluminescence. Gao et al.~\cite{gao2017interlayer} also recently applied this formula to study light emission in bilayer transition metal dichalcogenides. 
A similar equation was also employed by Cudazzo et al.~\cite{cudazzo2010strong} to investigate light emission in 2D materials, but it employed a prefactor that is incorrect for the 2D case. 
%
%
\subsection{One-dimensional materials}
\vspace{-10pt}
Excitons have been studied extensively in 1D materials, and first-principles calculations of exciton radiative lifetimes have been employed to investigate light emission in single-walled carbon nanotubes~\cite{avouris2008carbon,spataru2005theory}. 
Since defects and intertube interactions broaden and wash out the exciton spectrum, measuring exciton lifetimes is challenging in carbon nanotubes, 
and {\it ab initio} calculations have provided key microscopic insight into exciton recombination in carbon nanotubes~\cite{spataru2005theory}.
\\
\indent
In a 1D material, such as a nanotube or nanowire, the dimensionality constrains the exciton transition dipole to the direction of the material, which we take to be the $z$ direction. 
The transition dipole can then be written as $\bfp_S=p_{Sz}\hat{\bf z}$, and momentum conservation along the $z$ axis imposes a condition on the emitted photon wavevector, 
$\bfq\cdot\hat{\bf z}=\bfQ$, for the recombination of an exciton with momentum $\bfQ$ [see Fig. \ref{Fig:coordinate}(c)]. 
Using Fermi's Golden rule in Eq.~(\ref{eq Fermi Golden}), the exciton decay rate in a 1D material at zero temperature can be written as~\cite{spataru2005theory}:
\begin{equation}
\gamma_S^{\rm 1D}(\bfQ)=\gamma_S^{\rm 1D}(0)\cdot\frac{E_S(\bfQ)^2-\hbar^2c^2Q^2}{E_S(\bfQ)^2},
\end{equation}
where $\gamma_S^{\rm 1D}(0)=\frac{e^2p_{Sz}^2}{\e_0m^2\hbar c^2L_z}$ and $L_z$ is the
length of the system along the $z$ direction. The radiative recombination rate decreases monotonically with $Q$, and is zero when $Q_0=E_S(Q_0)/\hbar c$. 
Similar to the 2D case, $Q=Q_0$ is an upper limit to the exciton momentum for radiative recombination, and excitons with $Q>Q_0$ cannot recombine radiatively and emit light.
\\
\indent
The finite temperature radiative rate is computed using the thermal average in Eq.~(\ref{Eq:gammaST}). Assuming a parabolic exciton dispersion, 
the exciton radiative lifetime in a 1D material reads
\begin{equation}
\<\tau^{\rm 1D}_S\>(T)=
{\gamma_S^{\rm 1D}}(0)^{-1}
\times 
\frac{3}{4}
\left( \frac{\sqrt{2\pi M_S k_BT}}{E_S(0)/ c} \right).
\end{equation}
Using this equation, Spataru et al. obtained radiative lifetimes in carbon nanotubes in good agreement with experiment~\cite{spataru2005theory}.

%
%
\subsection{Atoms, molecules, and other isolated (0D) systems}
\vspace{-10pt}
We refer to an atom, molecule, quantum dot or other isolated light emitter as a 0D system [see Fig.~\ref{Fig:coordinate}(d)]. 
The approach presented here applies to both these isolated emitters and to atoms, ions or other single quantum emitters embedded in an isotropic material. 
Since there is no translation symmetry, the crystal momentum can be taken to be zero and ignored, and we keep only one quantum number to denote the discrete energy levels. Using these conventions, we write the exciton wavefunction in the Tamm-Dancoff approximation as
\begin{equation}
|S\>=\sum_{vc}A^S_{vc}|v\>_h|c\>_e
\end{equation} 
where $v$ and $c$ are quantum numbers associated with occupied and unoccupied orbitals, respectively. In general, when there are no symmetry constraints, the transition dipole is a complex vector, as in Eq.~(\ref{Eq:3D-dipole}). When the system is embedded in an isotropic material with dielectric constant $\e$ (for the 0D system in vacuum, one should set $\e=1$), Fermi's Golden rule gives the exciton recombination rate at zero temperature (see Appendix \ref{Appen:0D}):  
\begin{equation}
\label{Eq:0D-rate}
\gs^{\rm 0D}=\frac{\sqrt{\e}e^2p_S^2E_S}{3\pi\e_0m^2c^3\hbar^2}.
\end{equation} %
In CGS units, in which $\e_0 = 1/4\pi$, we recover the known result $\gs^{\rm 0D} \propto 4/3\, p_S^2E_S $ for the radiative rate of an isolated emitter or a defect embedded in a crystal~\cite{Dexter}, which is also known as the Einstein $A$ coefficient~\cite{Einstein} in the thermodynamic treatment of light emission. 
While light is quantized in our approach, we obtain the same formula as in Dexter's work in Ref.~\cite{Dexter}, where radiation is treated classically.
Due to the absence of crystal momentum for an isolated emitter, all the excitons satisfying the selection rules with nonzero transition dipole can undergo an optical transition and emit a photon. 
At finite temperature, since there is no momentum, we take a thermal average only over different exciton states [using Eq.~(\ref{Eq:3D gamma eff})], and obtain for the effective radiative recombination rate:
\begin{eqnarray}
\label{Eq:thermal 0D-rate}
\<\gamma^{\rm 0D}(T)\>_{\rm eff}=
\frac{\sqrt{\e}e^2}{3\pi\e_0m^2c^3\hbar^2}
\frac{\sum_S p_S^2E_Se^{-E_S/\kbt}}{\sum_S e^{-E_S/\kbt}}.
\end{eqnarray}

%
%
\section{Numerical calculations\label{Sect:data}}
\vspace{-10pt}
To our knowledge, there are no examples in the literature of {\it ab initio} calculations of radiative lifetimes in bulk crystals and 0D isolated systems within the BSE framework. 
We apply our approach to compute  from first principles the exciton radiative lifetimes in a bulk isotropic crystal of GaAs and in several small organic molecules in the gas phase. 
%
%
\begin{figure}[t]
\centering
\includegraphics[scale=0.25]{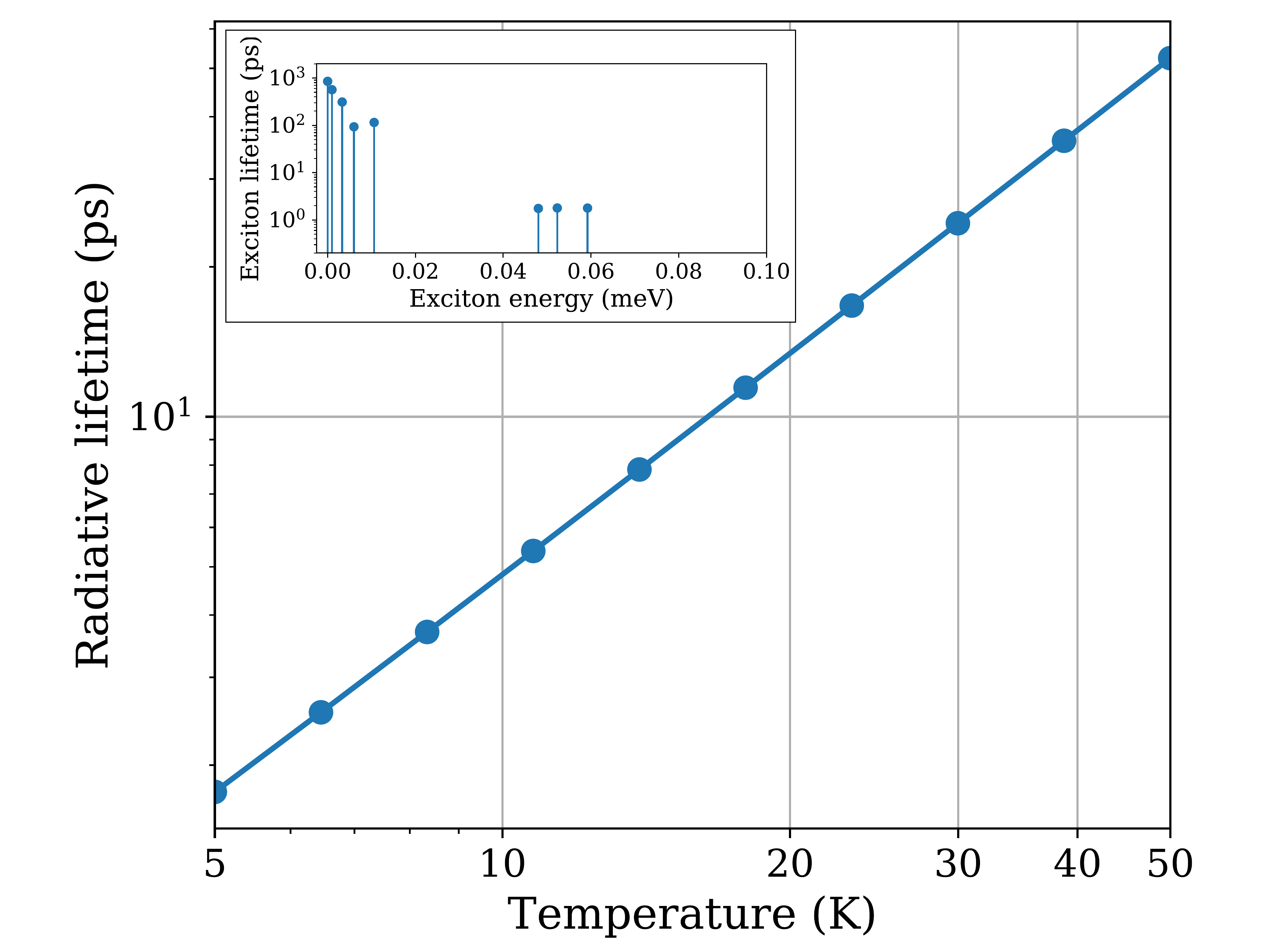}
\caption{Computed radiative lifetimes in a GaAs crystal, shown as a function of temperature up to 50 K. The lifetimes are obtained using the thermal average in Eq.~(\ref{Eq:3D gamma eff}).  
The inset shows the excitons contributing to the thermal average along with their individual lifetimes at 10 K. In the inset, the zero of the energy axis is taken to be the lowest exciton energy.}
\label{Fig:3D}
\end{figure}
To compute the radiative lifetimes, we first carry out density functional theory (DFT) calculations with the {\sc Quantum Espresso} code \cite{giannozzi2009quantum}, 
using normconserving pseudopotentials~\cite{NormCon1991} and a plane-wave basis set. 
We then carry out GW-BSE calculations with the {\sc Yambo} code~\cite{marini2009yambo}, and compute the radiative lifetimes by post-processing the BSE results.\\
\indent
For GaAs, we perform DFT calculations on the relaxed zincblende structure, employing the PBEsol exchange-correlation functional~\cite{PBEsol}. 
We use fully relativistic normconserving pseudopotentials generated with Pseudo Dojo~\cite{Dojo}, and include spin-orbit coupling in all calculations. 
The BSE is solved on a $30\times30\times30$ $\bfk$-point grid with a rigid scissor shift applied to the DFT band structure to match the experimental band gap \cite{levinshtein1996gaas}.  
We use a 6 Ry cutoff for the statically screened Coulomb interaction and the highest 4 valence bands and lowest 2 conduction bands to converge the low-energy excitons. 
In the radiative lifetime calculations, 
we use experimental values for the static dielectric constant and effective masses~\cite{levinshtein1996gaas,dargys1994handbook} to remove a possible source of error. 
%
%
Due to the light electron mass, which leads to a steep conduction band valley, fully converging the radiative lifetimes in GaAs requires very fine Brillouin zone grids with a large computational cost. 
Using a double-grid technique, Kammerlander et al.~\cite{Kammerlander} have shown that a $40\times40\times40$ $\bfk$-point grid is sufficient to converge the BSE absorption spectrum. 
Since the radiative lifetimes depend on the energy and transition dipole of the lowest-energy excitons, the $\bfk$-point convergence of the lifetimes is similar to that of the absorption spectrum. 
Using the  standard BSE (without the double-grid technique of Ref.~\onlinecite{Kammerlander}), the finest grid we were able to reach is a $30\times30\times30$ $\bfk$-point grid, 
which took over $10,000$ CPU cores to compute. Since our grid is close to the fully converged $40\times40\times40$ $\bfk$-point grid, 
it allows us to obtain results close to convergence. 
Further refinement of the GaAs radiative lifetimes given here may be possible by using finer grids, though their computational cost is at present prohibitive.
\\
\indent
For the gas-phase organic molecules, we use the experimental structure in all cases, and carry out all calculations at the $\Gamma$-point only. 
The DFT calculations employ the PBE exchange-correlation functional and a 90 Ry kinetic energy cutoff. 
In the BSE calculations, we use a 7 Ry cutoff for the statically screened Coulomb interaction, together with GW-corrected electron energy levels and up to 180 empty states to accurately converge the low-energy excitons. 
We employ cubic simulation cells with sizes between 33 $-$ 38 Bohr and use a truncated Coulomb interaction.\\
\indent
The computed radiative lifetimes in GaAs as a function of temperature are shown in Fig.~\ref{Fig:3D}.  
They are obtained as the thermal average in Eq.~(\ref{Eq:3D gamma eff}) of the BSE exciton radiative rates for a bulk isotropic crystal in Eq.~(\ref{Eq:gammaT}). 
The inset of Fig.~\ref{Fig:3D} shows the low-energy excitons contributing to this thermal average; the lowest 5 excitons are dark and associated with spin-forbidden transitions, and the 3 bright excitons at a slightly higher energy also contribute to the average. The dark states increase the average radiative lifetime by an order of magnitude compared to the average lifetime of the bright excitons alone. 
\\
\indent
The computed BSE radiative lifetimes are of order 1$-$50 ps below 50 K, and exhibit the $T^{3/2}$ trend expected for bulk crystals at low temperature~\cite{im1997radiative}. 
Comparing these results with experiment is not simple. In GaAs, the radiative processes are known to be affected by the coupling of excitons with phonons and free electron-hole pairs, 
resulting in an intricate nonequilibrium dynamics that is still the subject of debate~\cite{Foxon1987, GaAsExp1998, GaAsExp2016}. 
The interaction with phonons is particularly important in GaAs, where exciton-phonon scattering is thought to provide the momentum needed by excitons to transition toward the radiative region~\cite{Foxon1987,GaAsExp1998}.  
For this reason, the photoluminescence decay is expected to be much slower than the intrinsic exciton radiative lifetimes computed here. 
Consistent with this view, the measured photoluminescence decay times are a few ns at low temperatures~\cite{Foxon1987, GaAsExp1998,GaAsExp2016}, 
while our computed radiative lifetimes are a few ps in the same temperature range. 
This result confirms that the long lifetimes observed in GaAs by measuring the photoluminescence decay are the result of nonequilibrium exciton dynamics rather than an intrinsic exciton lifetime.  
Future work will investigate the coupled nonequilibrium dynamics of excitons and phonons, which will enable quantitative comparisons with photoluminescence data.\\
\indent
\begin{figure}[!t]
\centering
\includegraphics[scale=0.22]{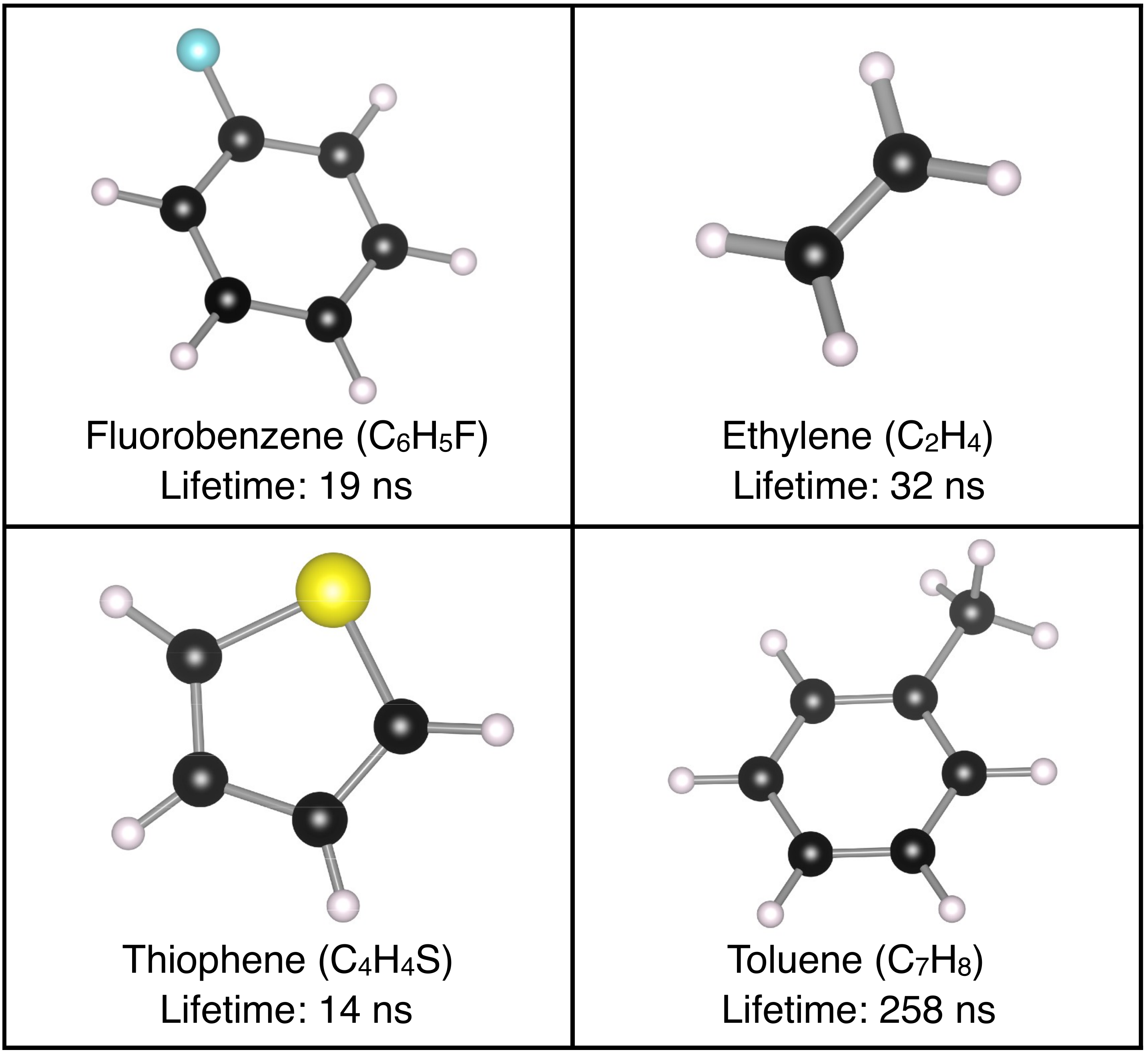}
\caption{The four molecules studied in this work $-$ fluorobenzene, ethylene, thiophene and toluene $-$ and their computed exciton radiative lifetimes.} 
\label{FIg:0D}
\end{figure}
\begin{table}[!b]
\begin{tabular}{|M{2.3cm}|M{1.8cm}M{2cm}M{2cm}N|}  
  \hline 
  & & & &\\[-3pt]
  & $\tau~({\rm ns})$ &$\tau_{\rm exp}~({\rm ns}) $& $p_S^2$ ($\times 10^6$) & \\[7pt]  
  \hline
& & & &\\[-2pt]
Fluorobenzene    & 19 & $12-23$ &7.54  &\\[7pt] 
Toluene     & 258 & 185 &0.59 &\\[7pt] 
Ethylene  & 32 & $-$ &4.59&\\[7pt] 
Thiophene & 14 &  $-$& 11.5 &\\[7pt] 
\hline
\end{tabular}
\caption{Comparison of our computed radiative lifetimes, $\tau=(\gamma^{\rm 0D})^{-1}$ obtained using Eq.~(\ref{Eq:0D-rate}), 
with experimental data from Ref.~\cite{breuer1972fluorescence}. For fluorobenzene, the experimental lifetime is given as a range, 
which is obtained by combining data from Refs.~\cite{berlman2012handbook,nakamura1970fluorescence} and the quantum yield from Ref.~\cite{al1971quenching}. 
The symbol \lq\lq$-$" means that we could not find experimental data. The square transition dipoles are provided in last column in atomic units.}
\label{Table:0D}
\end{table}
We additionally compute the radiative lifetimes in four small organic molecules $-$ fluorobenzene, ethylene, thiophene and toluene (see Fig.~\ref{FIg:0D}). 
We compare our results with available experimental data \cite{breuer1972fluorescence} in Table~\ref{Table:0D}. 
The predicted exciton lifetimes are within a factor of 2 of the measured values, and thus in very good agreement with experiment. 
We find radiative lifetime of order 10$-$30 ns in fluorobenzene, ethylene and thiophene; the lifetime in toluene is significantly longer, roughly 250 ns and thus an order of magnitude longer than in the other molecules. To explain this trend, we show in Table~\ref{Table:0D} that the square of the transition dipole of the lowest bright exciton is an order of magnitude smaller in toluene than in the other molecules; since the radiative lifetime is inversely proportional to the square dipole [see Eq.~(\ref{Eq:0D-rate})], the weaker transition dipole explains the longer lifetime in toluene. 
The simple intuition is that in molecules like toluene with small transition dipoles the electron and hole wavefunctions have a small overlap. Future work will attempt to correlate the lifetimes and dipoles with the molecular structure and exciton wavefunctions in a wider range of molecular structures.\\
\indent
One factor contributing to the small discrepancy with experiment is that we compute the radiative rate using the ground state molecular structure rather than its
excited-state counterpart. 
The molecular structure typically relaxes in the excited state from which light is emitted \cite{energy-derivatives}, 
leading to the so-called Stokes shift $-$ a redshift between the absorption onset and emission energy $-$ 
which can be sizable in small organic molecules. 
Since the exciton energies and transition dipoles are modified by the structural relaxation, 
these effects are expected to account for at least part of the small discrepancy between our computed radiative lifetimes and the experimental values. 
Future work will employ recently developed methods to relax the molecular structure in the excited state~\cite{ismail2003excited,hutter2003excited} and investigate how the structure impacts the radiative lifetimes.\\
\indent 
The formulas we obtain can be applied using transition dipole matrix elements that may or may not include electron-hole interactions. Yet, correctly treating excitons through the BSE is essential whenever the exciton binding energy is sizeable and whenever the low-energy optical transitions are excitonic in nature, which is the case in most molecules and nanomaterials, and in crystals with a large band gap or low dielectric screening. For example, for the gas-phase molecules, in which the independent-particle picture fails altogether to describe the optical excitations due to the large exciton binding energy, neglecting the electron-hole interaction leads to large errors in the optical spectra and also in the lifetimes. In toluene, the lowest bright exciton is the 27th eigenstate of the BSE Hamiltonian in order of increasing energy; it has an energy of 4.92 eV and a lifetime of 258 ns (versus an experimental value of 185 ns). By contrast, the 27th transition in the independent-particle Hamiltonian gives a lifetime of 83,000 seconds; the lowest bright transition in the independent-particle picture has an energy of 8.2 eV and a lifetime of 350 ns. Therefore, it is clear that the independent-particle approximation for radiative lifetimes in molecules gives large and uncontrolled errors, and that correctly treating the excitons with the BSE is essential for computing radiative lifetimes in molecules. Similar comparisons for bulk semiconductors will be discussed in detail elsewhere.

\section{Conclusions \label{conclusion}}
\vspace{-10pt}
We presented a general approach based on DFT and the {\it ab initio} BSE method to compute the radiative lifetimes in bulk crystals, 2D and 1D materials, and in 0D or isolated systems such as a molecule, quantum dot or single quantum emitter. 
%
%
Diagonalizing the BSE Hamiltonian in transition space is computationally expensive. Since the size of the transition space scales quadratically with the number of atoms $N$, 
a standard diagonalization of the BSE Hamiltonian will scale as $ N^6$. Yet, once its eigenvalues and eigenvectors are obtained, the radiative lifetimes calculations shown here only add a small computational overhead. 
The temperature dependence of the exciton radiative lifetime at low temperature is predicted to be proportional to $T^{3/2}$ in bulk, $T$ in 2D and $T^{1/2}$ in 1D materials.   
The bulk crystal treatment is applied to a GaAs crystal, where our computed intrinsic radiative lifetimes are shorter than the values measured by photoluminescence, 
but consistent with their interpretation in terms of nonequilibrium dynamics of excitons coupled to phonons and free carriers. 
Our approach for isolated emitters is applied to small organic molecules in the gas phase, 
giving computed radiative lifetimes in good agreement with experiment, up to corrections due to structural relaxation in the excited state. 
Our work provides a framework for predicting the \textit{intrinsic} exciton radiative lifetimes in materials with any dimensionality. 
Since the BSE is considered a gold standard for computing optical absorption and excitons~\cite{blase}, it is expected to also provide accurate results for radiative processes and light emission. 
These calculations can provide a benchmark for materials in which extrinsic effects due to impurities or interfaces dominate the ultrafast dynamics.  
They can also guide the interpretation of ultrafast spectroscopy measurements and the discovery of new quantum emitters with long radiative lifetimes.
\acknowledgements
\vspace{-10pt}
The authors thank Davide Sangalli for fruitful discussions. 
This work was partially supported by the Department of Energy under Grant No.~DE-SC0019166, which provided for theory and method development, 
and by the National Science Foundation under Grant No.~ACI-1642443, which provided for code development. 
M. P. acknowledges the Tor Vergata University for financial support through the mission sustainability project 2DUTOPI.  
This research used resources of the National Energy Research Scientific Computing Center, a DOE Office of Science User
Facility supported by the Office of Science of the U.S. Department of Energy under Contract No. DE-AC02-05CH11231.\\

\appendix
\section{Second quantization of light in anisotropic materials \label{Appen:photon quant}}
\vspace{-10pt}
Starting from Maxwell's equations in a material:
\begin{eqnarray}
\bsn\cdot \bfD=0~,~&&\frac{\partial \bfD}{\partial t}=\bsn\times \bfH~~;~\bfD=\e_0\bser \bfE ~,\nn\\
\bsn\cdot \bfB=0~,~&&-\frac{\partial \bfB}{\partial t}=\bsn\times \bfE~~;~\bfB=\mu_0\bfH~,
\end{eqnarray}
where $\e_0$ is the vacuum permittivity and $\mu_0$ the vacuum susceptibility, we define the vector potential $\bfA$ and the scalar potential $\Phi$:
\begin{equation}
\bfB=\bsn\times \bfA,~~\bfE=-\bsn\Phi-\frac{\partial \bfA}{\partial t}.
\end{equation}
We adopt a generalized Coulomb gauge, in which:
\begin{equation}
\Phi=0,~~~ \bsn\cdot\(\e_0\bser \bfE\)=0\,,
\end{equation}
and write the equation of motion for $\bfA$ as
\begin{equation}
\label{eq-EOM_A}
-\mu_0\e_0\bser\frac{\partial^2 
\bfA}{\partial t^2 }=\bsn\times(\bsn \times \bfA)=\bsn(\bsn\cdot \bfA)-\bsn^2\bfA.
\end{equation}
From Eq.~(\ref{eq-EOM_A}), we construct the Lagrangian
\begin{eqnarray}
\mathcal{L}&=&\frac{1}{2}\int d\bfr \[\e_0\bfE^T(\bfr)\bser\bfE(\bfr)-\frac{\bfB(\bfr)^2}{\mu_0}\]\nn\\
&=&
\frac{1}{2}\int d\bfr \[\e_0\dot { \bfA} ^T(\bfr)\bser\dot { \bfA}(\bfr)-\frac{\(\bsn\times\bfA\)^2}{\mu_0}\].
\end{eqnarray}
The conjugate momentum of the vector potential is
\begin{equation}
\bsP (\bfr)=\frac{\delta\mathcal{L}}{\delta \dot{\bfA}(\bfr)}=\e_0\bser\dot{\bfA}(\bfr),
\end{equation}
and by performing a Legendre transformation, we write the Hamiltonian as
\begin{equation}
\mathcal{H}=\int d\bfr \,\bsP\,\dot{\bfA}-\mathcal{L}=\frac{1}{2}\int d\bfr
\[\frac{\bsP^T \bser^{-1}\bsP}{\e_0}+\frac{\(\bsn\times\bfA\)^2}{\mu_0}\].
\end{equation}
Note that the Hamiltonian for classical electromagnetic field in vacuum can be recovered by setting $\bser=\mathbf{I}$.\\
\indent
To define the creation and annihilation operators for second quantization, we follow the standard procedure and expand the vector potential in terms of its eigenmodes, which are labeled by the index $\lambda$:
\begin{equation}
\label{Eq:Art}
\bfA(\bfr,t)=\sum_{\l}q_\l\bff_\l(\bfr)e^{i\w_\l t},
\end{equation} 
where $q_\l$ are constants representing the amplitudes and $\bff_\l(\bfr)$ satisfy
\begin{equation} \label{solve-omega}
\w_\l^2\mu_0\e_0\bser \bff_\l-\bsn\times\(\bsn\times \bff_\l\)=0.
\end{equation}
Since $\w_\l$ enters the equation as a square, both $+\w_\l$ and $-\w_\l$ can have the same $\bff_\l$ solution. However, since the vector potential is always real, we need $\bfA^\dagger=\bfA$, so that for each $q_\l\bff_\l(\bfr)e^{i\w_\l t}$ in Eq.~(\ref{Eq:Art}), there must exists a corresponding term $q'_\l\bff'_\l(\bfr)e^{-i\w_\l t}$ such that
\begin{equation}
q'_{\l}\bff'_{\l}(\bfr)=q^*_{\l}\bff^*_{\l}(\bfr).
\end{equation}
For convenience, we label this part of the solution as $-\l$:
\begin{equation}
q'_\l\bff'_\l(\bfr)e^{-i\w_\l t}=q_{-\l}\bff_{-\l}(\bfr)e^{i\w_{-\l} t}.
\end{equation}
To obtain an orthogonality condition for the solutions, we substitute $\bff_\l(\bfr)=\frac{\sqrt{\bser}^{-1}}{\sqrt{\mu_0\e_0}}\bfg_\l(\bfr)$ and get:
 \begin{equation}
\w_\l^2\bfg_\l-\frac{\sqrt{\bser}^{-1}}{\sqrt{\mu_0\e_0}}\bsn\times\(\bsn\times \frac{\sqrt{\bser}^{-1}}{\sqrt{\mu_0\e_0}}\bfg_\l\)=0.
\end{equation}
Now with $\w_\l^2$ as the eigenvalue, $\bfg_\l$ are eigenfunctions of a Hermitian operator and form an orthogonal solution set:
\begin{equation}
\int d\bfr ~\bfg^\dagger_\l(\bfr)\cdot \bfg_{\l'}(\bfr)=
\int d\bfr ~\mu_0\e_0\bff^\dagger_\l(\bfr)\bser \bff_{\l'}(\bfr)=
\delta_{\l,\l'}.
\end{equation}
In the following, we take plane waves as our eigenmodes, and label them by their polarization and momentum by substituting $\l \rightarrow ( \l,\bfq)$, $-\l \rightarrow (- \l,-\bfq)$. We also put
\begin{equation}
\bff_{\l\bfq}(\bfr)= \frac{\bfe_{\l\bfq}}{\sqrt{\mu_0\e_0}} e^{i\bfq\cdot\bfr}.
\end{equation}
The equation of motion becomes
\begin{equation}
\label{photon dispersion}
\w_{ \l\bfq}^2\mu_0\e_0\bser\bfe_{ \l\bfq}+\bfq\(\bfq\cdot \bfe_{ \l\bfq}\)-q^2\bfe_{ \l\bfq}=0,
\end{equation}
the orthogonality condition
\begin{equation}
\label{Eq:ortho}
\bfe^\dagger_{ \l\bfq}\bser\bfe_{ \l'\bfq}=\delta_{\l,\l'}\,\,,
\end{equation}
and the relation connecting $\l$ and $-\l$:
\begin{equation}
q^*_{\l\bfq}\bfe^*_{\l\bfq}=q_{-\l,-\bfq}\bfe_{-\l,-\bfq}.
\end{equation}
Then the vector potential can be written as
\begin{eqnarray}
&&\bfA(\bfr, t)=\sum_{\l\bfq}q_{\l\bfq}\frac{\bfe_{\l\bfq}}{\sqrt{\mu_0\e_0}} e^{i(\bfq\cdot\bfr+\w_{\l\bfq}t)}\nn\\
&&=c
\sum_{\l>0,\bfq}q_{\l\bfq}\bfe_{\l\bfq}
e^{i(\bfq\cdot\bfr+\w_{\l\bfq}t)}+q^*_{\l\bfq}\bfe^*_{\l\bfq}e^{-i(\bfq\cdot\bfr+\w_{\l\bfq}t)}\nn\\
\end{eqnarray}
and the conjugate momentum becomes:
\begin{equation}
\bsP(\bfr,t)=c\sum_{\l\bfq}iq_{\l\bfq}\w_{\l\bfq}\e_0\bser\bfe_{\l\bfq}e^{i(\bfq\cdot\bfr+\w_{\l\bfq} t)}.
\end{equation}
The Hamiltonian can be written as:
\begin{eqnarray}
\mathcal{H}&=&\e_0c^2V\sum_{\l\bfq}\w_{\l\bfq}^2q^*_{\l\bfq} q_{\l\bfq}\nn\\
&=&\e_0c^2 V\sum_{\l>0,\bfq}\w_{\l\bfq}^2(q^*_{\l\bfq} q_{\l\bfq}+q_{\l\bfq} q^*_{\l\bfq}),
\end{eqnarray}
where $V$ is the volume of the system. Finally, we can define creation and annihilation operators for $\l>0$:
\begin{equation}
\hat{a}_{\l\bfq}=c\sqrt{\frac{2V\w_{\l\bfq}\e_0}{\hbar  }}q_{\l\bfq},~~\[\hat{a}_{\l\bfq},\hat{a}^\dagger_{\l'\bfq'}\]=\delta_{\bfq,\bfq'}\delta_{\l,\l'}
\end{equation}
using which the vector potential operator becomes:
\begin{equation}
\bfA(\bfr, t)=\sum_{\l\bfq}\sqrt{\frac{\hbar}{2V\w_{\l\bfq}\e_0}}
\(\hat{a}_{\l\bfq}\bfe_{\l\bfq}e^{i\(\bfq\cdot\bfr+\w_{\l\bfq}t\)}+h.c.\)
\end{equation}
and the Hamiltonian:
\begin{equation}
\mathcal{H}=\sum_{\l\bfq}\hbar\w_{\l\bfq}\(\hat{a}^\dagger_{\l\bfq}\hat{a}_{\l\bfq}+\frac{1}{2}\).
\end{equation}
\section{Derivation of the radiative lifetime in isotropic 3D materials \label{Appen:3D}}
\vspace{-10pt}
We provide additional details for the derivation of the radiative recombination rate in isotropic 3D materials, Eq.~(\ref{Eq:gammaT}). In an isotropic bulk material with dielectric constant $\e$, 
the photon vector potential is given by Eq.~(\ref{Eq:vector potential}) with frequency $\w_{\bfq}=c \, |\bfq|/\sqrt{\e}$, and the IP and OOP polarization vectors are those in Eq.~(\ref{Eq:IP/OOP}). 
Due to momentum conservation, the summation over all possible final photon wavevectors in Eq.~(\ref{eq Fermi Golden}) is restricted to $\bfq=\bfQ$. As a result, we can write the radiative rate as
\begin{eqnarray}
&&\gamma_{S}^{\rm 3D, iso}(\bfQ)=
\frac{\pi e^2}{\e_0m^2VcQ\sqrt{\e}}
\left\{
\left|
-p_{Sx}\sin\varphi+p_{Sy}\cos\varphi
\right|^2_{\rm IP}
\right.\nn\\
&&~~~~~\left.
+
\left|
p_{Sx}\cos\theta\cos\varphi+
p_{Sy}\cos\theta\sin\varphi-
p_{Sz}\sin\theta
\right|_{\rm OOP}^2
\right\}\nn\\
&&~~~~~~\times
\delta\(E_S(\bfQ)-\frac{\hbar c Q}{\sqrt{\e}}\).
\end{eqnarray}
By setting $\cos\varphi=Q_x/Q_{xy}$, where $\bfQ_{xy}$ is the projection of $\bfQ$ onto the $xy$ plane, and $\cos\theta=Q_z/Q$, we obtain Eq.~(\ref{Eq:iso rate}). 
To obtain the radiative rate at finite temperature $T$, we plug Eq.~(\ref{Eq:iso rate}) into Eq.~(\ref{Eq:gammaST}) along with the parabolic dispersion in Eq.~(\ref{Eq:parabola dispersion}). 
The denominator, due to lack of angular dependence, reduces to a Gaussian integral of the kind $\int^{\infty}_0\!dx\, x^2 \exp(-x^2)=\sqrt{\pi}/4$, and gives
\begin{eqnarray}
&&\int dQ_xdQ_ydQ_z e^{-E_S(Q)/k_BT}\nn\\
&&~~= \int \!d\Omega  \int^\infty_0\!\! dQ\, Q^2 \,e^{\frac{-\hbar^2Q^2}{2M_Sk_BT}}=\(\frac{2\pi M_S k_BT}{\hbar^2} \)^{3/2}
\end{eqnarray}
where $d\Omega = \sin\theta d\theta d \varphi$ is the differential solid angle, and we leave out the factor $e^{-E_S(0)/k_BT}$, 
which is present both in the numerator and denominator and cancels out in the final result. 
For the numerator, we note that the exciton parabolic dispersion can be approximated as flat within the light cone, so that we can put $E_S(\bfQ)\approx E_S(0)$. As a result, we get 
\begin{widetext}
\begin{eqnarray}
&&\int \!dQ_xdQ_ydQ_z\, e^{-E_S(Q)/k_BT}\, \gamma_{S}^{\rm 3D, iso}(\bfQ)=\int \!d\Omega \int^\infty_0\!\! dQ\,\, Q^2 e^{-E_S(Q)/k_BT} \,\gamma_{S}^{\rm 3D, iso}(\bfQ)\nn\\
&&\approx
\frac{\pi e^2 }{\e_0m^2Vc\sqrt{\e}}
\int \!d\Omega \int\!dQ \,\,Q 
\left\{
\left|
-p_{Sx}\sin\varphi+p_{Sy}\cos\varphi
\right|^2\right.
\nn\\
&&\left.~~~~~~~~~~~~~~~
+
\left|
p_{Sx}\cos\theta\cos\varphi+
p_{Sy}\cos\theta\sin\varphi-
p_{Sz}\sin\theta
\right|^2
\right\}
\delta\!\(E_S(0)-\frac{\hbar c Q}{\sqrt{\e}}\)
\nn\\
&&=
\frac{\pi e^2 }{\e_0m^2Vc\sqrt{\e}}
\int \! d \varphi \,d\theta\, \sin\theta \int\!\!dQ \,\,Q
\left\{
|p_{Sx}|^2\sin^2\varphi+|p_{Sy}|^2\cos^2\varphi
\right.
\nn\\
&&\left.~~~~~~~~~~~~~~~
+
|p_{Sx}|^2\cos^2\theta\cos^2\varphi+
|p_{Sy}|^2\cos^2\theta\sin^2\varphi+
|p_{Sz}|^2\sin^2\theta
\right\}
\delta\!\(E_S(0)-\frac{\hbar c Q}{\sqrt{\e}}\)
\nn\\
&&=\frac{8\pi^2 e^2p_S^2 }{3\e_0m^2Vc\sqrt{\e}} \int \!dQ \,\,Q
\cdot\delta\!\(E_S(0)-\frac{\hbar c Q}{\sqrt{\e}}\)
=
\frac{8\pi^2\sqrt{ \e}e^2 p_S^2E_S(0)}{3\e_0\hbar^2 c^3m^2V}.
\end{eqnarray}
\end{widetext}
\newpage

After dividing the numerator by the denominator, we obtain
\begin{eqnarray}
\<\gamma_S^{\rm 3D, iso}\>(T)=
\frac{8\sqrt{\pi \e}\,e^2\,\hbar \,p_S^2}{3\,\e_0 m^2VE_S(0)^2}
\(\frac{E_S(0)^2}{2M_Sc^2k_BT}\)^{3/2}
\end{eqnarray}
namely the finite temperature radiative lifetime in Eq.~(\ref{Eq:gammaT}).
%
%
\section{Derivation of the radiative lifetime in 0D systems \label{Appen:0D}}
\vspace{-10pt}
We provide additional details for the derivation of the radiative recombination rate in 0D systems, Eq.~(\ref{Eq:0D-rate}). 
As discussed above, in the 0D case there is no constraint from momentum conservation on the emitted photon wavevector. Therefore, we replace the summation in Eq.~(\ref{eq Fermi Golden}) by an integration over the full momentum space and write 
\begin{equation}
\gs^{\rm 0D}=
\frac{\pi e^2}{\e_0m^2}
\sum_{\l}
\int\frac{d\Omega \, dq\, q^2}{(2\pi)^3}
\frac{\left|\bfe_{ \l\bfq}\cdot\bfp_S(\bfQ)\right|^2}{\w_{\l\bfq}}
\delta\( E_S-\hbar \w_{\l\bfq}\) 
\end{equation}
which comes from rewriting the summation along each cartesian component $\alpha$ as $\sum_{q_\alpha}=\int L_\alpha dq_\alpha/2\pi$, and $L_x L_y L_z=V$.
In the 0D case, we can apply the photon quantization solutions used in the 3D case, 
so that $\lambda=$ IP or OOP, and $\bfe_{ \l\bfq}$ are in the form of Eq.~(\ref{Eq:IP/OOP}) with $\w_{\l\bfq}=c|\bfq|/\sqrt{\e}$. 
Note that this approach applies both to isolated emitters, such as quantum dots and molecules, as well as to atoms, ions or other single quantum emitters embedded in an isotropic material. 
Combining these results, we can write 
\begin{eqnarray}
&&\gs^{\rm 0D}=\frac{\pi e^2}{8\e_0m^2\pi^3c\sqrt{\e}}\times\nn\\
&&~\int
\!\!d\varphi\, d\theta \sin\theta \!\int \!\!dq \,q \left\{
\left|
-p_{Sx}\sin\varphi+p_{Sy}\cos\varphi
\right|^2\right.
\nn\\
&&~~~\left.
+
\left|
p_{Sx}\cos\theta\cos\varphi+
p_{Sy}\cos\theta\sin\varphi-
p_{Sz}\sin\theta
\right|^2
\right\}\nn\\
&&\qquad\qquad\qquad\qquad\qquad\qquad\qquad\qquad\times
\delta\!\(E_S-\frac{\hbar c q}{\sqrt{\e}}\)\nn\\
\end{eqnarray}
and finally obtain Eq.~(\ref{Eq:0D-rate}),
\begin{equation}
\gs^{\rm 0D}=\frac{\sqrt{\e}e^2p_S^2E_S}{3\pi\e_0m^2c^3\hbar^2}.
\end{equation}

\vspace{40pt}
\section{Radiative lifetime of excitons with linear dispersion \label{Appen:non-analytic}}
\vspace{-10pt}
We provide an additional discussion for excitons with linear dispersion:
\begin{equation}
\label{Eq:non-analytic dispersion}
E_S(\bfQ)=E_S(0)+B|\bfQ|.
\end{equation} 
The radiative lifetime at finite temperature can still be derived using Eq.~(\ref{Eq:gammaST}). The linear exciton dispersion merely changes the phase-space integral in the denominator of Eq.~(\ref{Eq:gammaST}), 
leading to a simple extension of the treatment for parabolic exciton dispersion. The phase-space integral in $d$ dimensions can be written as:
\begin{equation}
I(d)=\int d\Omega_d\int d Q Q^{d-1} \exp\[\frac{-BQ}{k_BT}\]
\end{equation}
where $d$ is the dimensionality of the material and $\Omega_d$ is the $d-$dimensional differential solid angle. 
We obtain:
\begin{equation}
I(d)=\left\{
\begin{matrix}
8\pi(k_BT/B)^3~~&d=3&\\
2\pi(k_BT/B)^2~~&d=2&\\
2(k_BT/B)~~&d=1&
\end{matrix}
\right. .
\end{equation}
Using this result together with Eq.~(\ref{Eq:gammaST}), we obtain the radiative lifetimes in isotropic 3-, 2- and 1-dimensional materials with linear exciton dispersion:
\begin{equation}
\<\gamma_S^{\rm d,iso}\>_{\rm linear}(T)=
\left\{
\begin{matrix}
\frac{\pi \sqrt{\e}e^2 \hbar p_S^2}{3\e_0m^2 VE_S(0)^2}\(\frac{BE_S(0)}{ck_B\hbar T}\)^3~~&d=3&\\
\gamma^{\rm 2D}_S(0)\times \frac{2}{3}
\(\frac{BE_S(0)}{ck_B\hbar T}\)^2~~&d=2&\\
\gamma^{\rm 1D}_S(0)\times \frac{2}{3}
\frac{BE_S(0)}{ck_B\hbar T}~~&\,d=1,&
\end{matrix}
\right. 
\end{equation}
where $\gamma^{\rm 2D}_S(0)$ and $\gamma^{\rm 1D}_S(0)$ are the intrinsic radiative rates in 2D and 1D systems, respectively, which are defined in the main text and are independent of the exciton dispersion. 
The radiative lifetimes, $\<\tau_S \>= \< \gamma_S\>^{-1}$, for excitons with linear dispersion exhibit a stronger temperature dependence, $\<\tau_S^{\rm d,iso}\>_{\rm linear}(T) \propto T^d$, versus $\<\tau_S^{\rm d,iso}\>_{\rm parabolic}(T) \propto T^{d/2}$ for the parabolic exciton dispersion case. For 2D materials with linear exciton dispersions, which have been recently predicted, the radiative lifetimes are thus expected to follow a T$^2$ trend with temperature.
%
\bibliographystyle{apsrev4-1}
%

\end{document}